\documentclass{article}
\usepackage{dcase2016,amsmath,graphicx,url,times}
\usepackage{resizegather}
\usepackage{array}
\usepackage{cite}
\newcolumntype{?}{!{\vrule width 1.25pt}}

\title{Experiments on the DCASE Challenge 2016: Acoustic Scene Classification and Sound Event Detection in Real Life Recording}

\name{Benjamin Elizalde\textsuperscript{1}, Anurag Kumar\textsuperscript{1}, Ankit Shah\textsuperscript{2}, Rohan Badlani\textsuperscript{3}, 
Emmanuel Vincent\textsuperscript{4}, Bhiksha Raj\textsuperscript{1}, Ian Lane\textsuperscript{1}}
\address{\textsuperscript{1}Carnegie Mellon University, Pittsburgh, PA, USA, \textsuperscript{4}Inria, F-54600 Villers-l\`es-Nancy, France \\
\textsuperscript{2}NIT Surathkal, India,\textsuperscript{3}BITS, Pilani, India\\
     bmartin1,alnu@andrew.cmu.edu, rohan.badlani,ankit.tronix@gmail.com,\\              	emmanuel.vincent@inria.fr, bhiksha@cs.cmu.edu,lane@cmu.edu}

\begin{document}

\ninept
\maketitle

\begin{sloppy}

\begin{abstract}
In this paper we present our work on Task 1 Acoustic Scene Classification and Task 3 Sound Event Detection in Real Life Recordings. Among our experiments we have low-level and high-level features, classifier optimization and other heuristics specific to each task. Our performance for both tasks improved the baseline from DCASE: for Task 1 we achieved an overall accuracy of $\mathbf{78.9\%}$ compared to the baseline of $\mathbf{72.6\%}$ and for Task 3 we achieved a Segment-Based Error Rate of $\mathbf{0.76}$ compared to the baseline of $\mathbf{0.91}$.
\end{abstract}

\vspace{-0.1in}
\section{Introduction}
\label{intro}
\vspace{-0.1in}

Audio plays a critical role in understanding the environment around us. This makes audio content analysis research important for tasks related to multimedia \cite{cheng2012sri,jiang2015fast}, and human computer interaction~\cite{chu2006scene,janvier:hal-00768767} to mention a pair. However, unlike the field of computer vision which has a variety of standard publicly available datasets such as Imagenet, audio event/scene analysis lacks such large dataset. This makes it difficult to compare different approaches and establishing the state of art. The second iteration of DCASE~\cite{Mesaros2016_EUSIPCO}, occurring in 2016, offers an opportunity to compare approaches on a standard public dataset. This edition it includes four different tasks: acoustic scene classification, sound event detection-- real and synthetic audio, and audio tagging. 

The state-of-the-art of the previous DCASE challenge, for both acoustic scenes~\cite{roma2013recurrence,rakotomamonjy2015histogram,schroder2013use} and sound event detection~\cite{roma2013recurrence,schroder2013use,gemmeke2013exemplar}, attributed their success mainly to features and audio representations rather than classifiers. Hence, an important aspect in our work is to emphasize on classifier exploration along with features. In this paper we present our work performed on Task 1 and Task 3. We proposed a variety of methods for both tasks and we obtained significant improvement over the baseline methods. 


\vspace{-0.2in}
\section{Tasks and Data}
\vspace{-0.1in}
\label{sec:data}
The goal of Task 1, Acoustic Scene Classification, is to classify a test recording into one of predefined classes that characterizes the environment in which it was recorded — for example \textit{park}, \textit{home}, \textit{office}. TUT Acoustic Scenes 2016 dataset is used for this task. It consists of recordings from various acoustic scenes. For each recording location, a 3-5 minute long audio recording was captured. The original recordings were then split into 30-second segments for the challenge. There are 15 acoustic scenes for the task.

Task 3, Sound Event Detection in Real Life Recordings, evaluates performance of sound event detection in multi-source conditions similar to our everyday life. There is no control over the number of overlapping sound events at each time, not in the training nor in the audio data. TUT Sound Events 2016 dataset is used for Task 3, which consists of recordings from two acoustic scenes: \textit{Home} and \textit{Residential Area}. There are 18 selected sound event classes, 11 for Home and 7 for Residential Area.
\vspace{-0.15in}

\section{Task 1: Acoustic Scene Classification}
\vspace{-0.1in}
\label{sec:task1}
From machine learning perspective, we treated Task 1 as a multi-class classification problem. The first step is to use a suitable method for characterizing acoustic scenes in the audio segments. An effective approach for characterizing audio events is bag-of-audio-words based feature representation \cite{bow}, which is usually built over low-level features such as MFCCs. Acoustic scenes, however, are more complex mixtures of different audio events and a more robust representation is required. To obtain a more robust representation we use Gaussian Mixture Models (GMMs) for feature representations of audio segments. Broadly, we employed two high-level feature representations to represent audio scenes. On the classification front we used Support Vector Machines (SVMs) as our primary classifier and in combination with other classifiers. 
\vspace{-0.1in}
\subsection{Feature Representations}
\vspace{-0.1in}
Let $D$-dimensional MFCCs vectors for a recording be represented as $\vec{x}_t$, where $t=1\,\,$to$\,\,T$, $T$ is the total number of MFCCs vectors for the recording. The major idea behind both high-level feature representation is to capture the distribution of MFCCs vectors of a recording. We will refer to these features as $\vec{\alpha}$ and $\vec{\beta}$ features and the sub-types will be represented using appropriate subscripts and superscripts. 

The first step in obtaining high-level fixed dimensional feature representation for audio segments is to train a GMM on MFCC vectors of the training data. Let us represent this GMM by $\mathcal{G} = \{w_k,N(\vec{\mu}_k, \Sigma_k), k = 1 \,\,to \,\,M\}$, where $w_k$, $\vec{\mu}_k$ and $\Sigma_k$ are the mixture weight, mean and covariance parameters of the $k^{th}$ Gaussian in $\mathcal{G}$. We will assume diagonal covariance matrices for all Gaussians and $\vec{\sigma}_k$ will represent the diagonal vector of $\Sigma_k$. Given the MFCCs vectors $\vec{x}_t$ of a recording, we computed the probabilistic assignment of $\vec{x}_t$ to the $k^{th}$ Gaussian. These soft assignments are added over all $t$ to obtain the total mass of MFCCs vectors belonging to the $k^{th}$ Gaussian (Eq \ref{eq:addms}). Normalization by $T$ is used to remove the effect of the duration of recordings. 
\begin{equation}
\label{eq:addms}
\resizebox{0.90\columnwidth}{!}{$ Pr(k | \vec{x}_{t}) =  \frac{w_{k}N(\vec{x}_{t} ; \vec{\mu}_k, \Sigma_k)}{\sum\limits_{j=1}^M w_jN(\vec{x}_{t} ; \vec{\mu}_k, \Sigma_k)},P(k) =  \frac{1}{T}\sum\limits_{i=1}^T Pr(k | \vec{x}_{t})$}
\vspace{-0.05in}
\end{equation}
The soft count histogram features referred to as $\vec{\alpha}$ is, $\vec{\alpha}^M=[P(1),..P(k)..P(M)]^T$. $\vec{\alpha}^M$ is an $M$-dimensional feature representation for a given recording. It captures how the MFCC vectors of a recording are distributed across the Guassians in $\mathcal{G}$. $\vec{\alpha}^M$ is normalized to sum to $1$ before using it for classifier training. 

The next feature ($\vec{\beta}$), also based on the GMM $\mathcal{G}$, tries to capture the actual distribution of the MFCC vectors of a recording. This is done by adapting the parameters of $\mathcal{G}$ to the MFCC vectors of the recording. We employ maximum {\em a posteriori} (MAP) estimation to for the adaptation \cite{gauvain1994} \cite{bimbot}. Parameter adaptation for $k^{th}$ Gaussian follows the following steps. First we compute, 
\begin{equation}
\resizebox{1\columnwidth}{!}{$n_{k}=\sum\limits_{t=1}^T Pr(k | \vec{x}_{t}),\,\,\,E_{k}(\vec{x})=\frac{1}{n_{k}}\sum\limits_{t=1}^T Pr(k | \vec{x_{t}})\vec{x}_{t},\,\,E_{k}(\vec{x}^2)=\frac{1}{n_{k}}\sum\limits_{t=1}^T Pr(k | \vec{x_{t}})\vec{x}_{t}^2$}
\end{equation}
Finally, the updated mean and variances are obtained as 
\begin{align}
\hat{\vec{\mu}}_k=&\frac{n_k}{n_k+r}E_{k}(\vec{x})+\frac{r}{n_k+r}\vec{\mu}_k \\
\hat{\vec{\sigma}}_k=&\frac{n_k}{n_k+r}E_{k}(\vec{x}^2)+\frac{r}{n_k+r}(\vec{\sigma}_k^2+\vec{\mu}_k^2) - \hat{\vec{\mu}}_k^2
\end{align}
The relevance factor $r$ controls the effect of the original parameters on the new estimates. We obtain $2$ different feature representation using the adapted means ($\hat{\vec{\mu}}_k$) and variances ($\hat{\vec{\sigma}}_k$). The first one denoted by $\vec{\beta}^M$ is an $M \times D$ dimensional feature obtained by concatenating the adapted means $\hat{\vec{\mu}}_k$ for all $k$, that is $\vec{\beta}^M=[\hat{\vec{\mu}}_1^T,...\hat{\vec{\mu}}_K^T]^T$. In the second $\vec{\beta}$ features adapted $\hat{\vec{\sigma}}_k$ are concatenated along with $\hat{\vec{\mu}}_k$ to obtain a $2 \times M \times D$ dimensional features. This form of $\vec{\beta}$ features are denoted by $\vec{\beta}^M_{\sigma}$. 
\vspace{-0.15in}

\subsection{Classification}
\vspace{-0.1in}
Once the feature representation for audio segments have been obtained, Task 1 essentially becomes a multi-class classification problem. Our primary classifiers are SVMs where we explore a variety of kernels. For the $\vec{\beta}$ features, we use Linear Kernel (LK) and RBF Kernel (RK). For soft-count histogram $\vec{\alpha}$ features we explore a panoply of kernels. Along with LK and RK we explored the following kernels.
\begin{itemize}
\item Exponential $\chi^2$ Distance (ECK): the kernel is computed as $K(\vec{x},\vec{y})=\exp^{-\gamma D(\vec{x},\vec{y})}$, where $D(\vec{x},\vec{y}) = \sum_i (x_i-y_i)^2/(x_i+y_i)$ is $\chi^2$ distance. 
\item $\chi^2$ Kernel (CK): In this case $K(\vec{x},\vec{y})=\sum_i \frac{2 x_i y_i}{x_i+y_i}$
\item Intersection Kernel (IK): $K(\vec{x},\vec{y})=\sum_i \min(x_i,y_i)$
\item Exponential Hellinger Distance Kernel (EHK): $K(\vec{x},\vec{y}) = \exp^{-\gamma D(\vec{x},\vec{y})}$ where $D(\vec{x},\vec{y}) =  \sum_i (\sqrt{x_i} - \sqrt{y_i})^2$
\item Hellinger Kernel (HK): $(\vec{x},\vec{y}) = \sum_i \sqrt{x_i y_i}$
\end{itemize}  
The details of these kernels can be found in \cite{zhang2007local,vedaldi2012efficient,li2013sign}. For kernels where $\gamma$ term appears, the optimal value of $\gamma$ value can be obtained by cross validation over training data. However, setting $\gamma$ equal to the inverse of average distance $D(\vec{x},\vec{y})$ between training data points works well in general as well. We use \cite{LIBSVM}\cite{fanliblinear} for SVM implementation.

Finally, we have a classifier fusion step where we combined the output of the different classifiers. We combined multiple classifiers by taking prediction vote from each classifier and the final predicted class is the one which gets the maximum vote. We call it the \emph{Fused Classifier} and we observed that the fused classifier can give significant improvement for several acoustic scenes. 
\vspace{-0.15in}

\subsection{Results}
\vspace{-0.1in}
Our experimental setup with the folds structure, is same as the one provided by DCASE. We extracted 20 dimensional MFCC features using $30ms$ window and $50\%$ overlap. MFCCs are augmented with their delta and acceleration features. For our final feature representation we experimented with $4$ different values of GMM component size $M$, $64,128,256$ and $512$. The relevance factor $r$ for $\vec{\beta}$ is set to $20$. Due to space constraints we cannot present fold-and-scene specific results for all cases and hence overall accuracy for all $4$ folds is shown. Table \ref{tab:task11} shows overall accuracy results for different cases. The accuracy for the MFCC-GMM \emph{baseline} method provided in the challenge is $72.6\%$. 

We can observe from Table \ref{tab:task11} that $\vec{\alpha}$ features in general do not perform better than the baseline method for any SVM kernel. However, $\vec{\beta}$ features clearly outperformed baseline method. In the best case, with $M=128$ and $\vec{\beta}_\sigma^M$ our method outperformed the baseline by an absolute $5\%$. 

Table \ref{tab:task12} shows results for the fused classifiers. For the fusion step we did not consider classifiers built over $\vec{\alpha}$ since these classifiers are inferior compared to those using $\vec{\beta}$ features. We can observe that our proposed method beats the baseline method by an absolute $\mathbf{6.3\%}$. Moreover, for scenes such as \emph{Park, Train, Library} where the baseline method gives very poor results, we improved the accuracy by an absolute $\mathbf{16-30\%}$. We also obtained superior overall accuracy on all folds which suggests that our proposed method is fairly robust. This is further supported by the fact that on DCASE evaluation set, We achieved an overall accuracy of $\mathbf{85.9\%}$. 
\begin{table}[t]
\centering
\caption{Task 1 Accuracy for different cases (Single Classifier)}
\label{tab:task11}
\resizebox*{1.0\columnwidth}{!}{
\begin{tabular}{|c|c|c|c|c|c|c|c|c|c|c|c|}
\hline  
 &\multicolumn{7}{c|}{$\vec{\alpha}^M$} &  \multicolumn{2}{c|}{$\vec{\beta}^M$} & \multicolumn{2}{c|}{$\vec{\beta}^M_{\sigma}$}\\ 
\cline{2-12}
$M$&LK&RK&ECK&CK&IK&EHK&HK&LK&RK&LK&RK\\
\hline 
64&62.8&60.6&66.2&66.3&66.0&64.7&65.3&76.8&76.6&75.5&$\mathbf{76.7}$\\
\hline
128&63.6&62.3&67.5&67.1&66.4&67.4&66.5&76.5&75.3&77.5&$\mathbf{77.5}$\\
\hline
256&63.9&63.9&67.3&67.8&66.5&68.7&67.7&76.5&71.9&$\mathbf{76.6}$&75.9\\
\hline
512&65.0&62.9&67.8&67.8&67.1&68.9&69.3&$\mathbf{76.4}$&72.2&76.2&75.9\\
\hline
\end{tabular}
}
\vspace{-0.20in}
\end{table}
\begin{table}[t]
\centering
\caption{Overall Task 1 Accuracy (Fused Classifier)}
\label{tab:task12}
\resizebox*{1.0\columnwidth}{!}{
\begin{tabular}{|c|c|c|c|c|c?c|c|c|c|c|}
\hline  
 &\multicolumn{5}{c?}{\textbf{Baseline}} &  \multicolumn{5}{c|}{\textbf{Proposed}}\\ 
\cline{2-11}
Scene&Fold 1&Fold 2&Fold 3&Fold 4& \textbf{Avg.} & Fold 1&Fold 2&Fold 3&Fold 4 & \textbf{Avg.}\\
\hline 
Beach&84.2&66.7&78.9&47.4&69.3&100&71.4&89.5&52.6&78.4\\
\hline
Bus&68.4&65.0&100&85.0&79.6&68.4&50.0&100&95.0&78.4\\
\hline
Cafe/Restaurant&66.7&94.7&71.4&100&83.2&88.9&63.2&76.2&95.0&80.8\\
\hline
Car&70.0&89.5&89.5&100&87.3&80.0&100&100&100&95.0\\
\hline
City Center&83.3&73.7&89.5&95.5&85.5&88.9&84.2&100&95.5&92.1\\
\hline
Forest Path&57.1&100&66.7&100&81.0&81.0&100&100&100&95.2\\
\hline
Grocery Store&52.6&81.0&89.5&36.8&65&89.5&81.0&94.7&84.2&87.3\\
\hline
Home&100&55.6&95.0&77.8&82.1&100&61.1&80.0&44.4&71.4\\
\hline
Library&47.6&38.9&15.0&100&50.4&47.6&33.3&85.0&100&66.5\\
\hline
Metro Station&84.2&94.4&100&100&94.7&94.7&94.4&100&100&97.3\\
\hline
Office&100&100&94.4&100&98.6&78.9&100&72.2&83.3&83.6\\
\hline
Park&10.0&5.6&0&40.0&13.9&65.0&33.3&50.0&30.0&44.6\\
\hline
Residential&78.9&47.6&100&84.2&77.7&84.2&42.9&94.7&57.9&69.9\\
\hline
Train&16.7&31.6&30.4&61.1&34.9&50.0&63.2&34.8&88.9&59.2\\
\hline
Tram&88.9&88.9&63.6&100&85.3&83.3&88.9&63.6&100&84.0\\
\hline
\textbf{Overall}&\textbf{67.2}&\textbf{68.9}&\textbf{72.3}&\textbf{81.9}&\textbf{72.6}&\textbf{80.0}&\textbf{71.1}&\textbf{82.7}&\textbf{81.8}&\textbf{78.9}\\
\hline
\end{tabular}
}
\vspace{-0.10in}
\end{table}

\vspace{-0.1in}
\section{Task 3: Sound Event Detection in Real Life Recordings} 
\label{sec:task3}
\vspace{-0.1in}
\begin{table}[t]
\small
\centering
\resizebox*{1.0\columnwidth}{!}{
  \begin{tabular}{ | l | c | c | }
    \hline
    \textbf{Feature Type} & \textbf{Accuracy}\% & \textbf{Classifier}  \\ \hline
    MFCCs & 67.7 & Logistic Regression \\\hline
    GBFB & 52.4 & Gradient Boosting  \\\hline
    SGBFB & 61.5 & Gradient Boosting   \\\hline
    Scatnet & 62.1 & Random Forest   \\\hline
    Stacked & 66.68 & Random Forest   \\\hline
    Stacked + PCA& 66.06 & Random Forest   \\\hline
  \end{tabular}
  }
\caption{Sound-event classification accuracy for different feature types using the 18 sounds and the 75 training - 25 testing ratio. Stacked included normalized MFCCs +SGBFB +Scatnet.}
\label{featureTable}
\vspace{-1em}
\end{table}

Detection of sound events in scenes and long recordings have been treated as a multi-class classification problem before in\cite{elizaldeaudio,elizalde2013lost,Elizaldeacm} where a classifier is trained with the sound segments. For testing, the classifier outputs segment/frame-level predictions for all the classes. In order to follow a similar approach, first we wanted to analyze features' performance for sound events regardless of the scene. This way, we could have an intuition of performance on the harder scenario of Task 3 where not every segment of the scene corresponds to a labeled sound event. 
 
\vspace{-0.1in}
\subsection{Features and Classifiers Optimization}
\label{sec:features}
\vspace{-0.1in}

For the features we tried the conventional MFCCs with standard parameters such as 12 coefficients plus energy, delta and double delta for a total of 39 dimensions. Moreover, we explored three features addressing the time-frequency acoustic characteristics. The Gabor Filter Bank (GBFB) in ~\cite{schadler2012spectro} have 2D-filters arranged by spectral and temporal modulation frequencies in a filter bank. The Separable Gabor filter bank (SGBFB) features extract spectro-temporal patterns with two separate 1D GBFBs, a spectral one and a temporal one. This approach reduces the complexity of the spectro-temporal feature extraction and further improves robustness as demonstrated in ~\cite{schadler2015separable}. Both features have the default parameters from the toolbox\footnote{http://www.uni-oldenburg.de/mediphysik-akustik/mediphysik/downloads/gabor-filter-bank-features/} for a total dimension of 1,020 each. The Scatnet~\cite{sifre2013rotation} features are generated by a scattering architecture which computes invariants to translations, rotations, scaling and deformations, while keeping enough discriminative information. It can be interpreted as a deep convolution network, where convolutions are performed along spatial, rotation and scaling variables. As opposed to standard convolution networks, the filters are not learned but are scaled and rotated wavelets. The features were extracted with a toolbox\footnote{http://www.di.ens.fr/data/software/scatnet/} using 0.25 second segments. The dimensionality of the three Scatnet components are 2, 84, 435 for a total of 521. Additionally, we included the normalized (mean and variance) Stacked (MFCCs+ SGBFB+ Scatnet) with PCA and also the normalized (mean and variance) Stacked without PCA. For the PCA we used Scikit's~\cite{scikit-learn} and used the full dimensionality of 1,580 as the number of input components and the resultant automatic reduction was 909 dimensions. For all the feature types and for the sake of avoiding the length variability of the temporal dimension, we averaged the vectors across time to end up with one single vector per sound event file. 

Then, for the classifiers we considered Tpot~\cite{Olson2016EvoBIO}, built on top of Scikit~\cite{scikit-learn}, which is a Python tool that automatically creates and optimizes machine learning pipelines using genetic programming. This toolbox (version 4) considers 12 classifiers such as Decision Tree, Random Forest, Xtreme Gradient Boosting, SVMs, K-Neighbors and Logistic Regression. The main Tpot parameter is ``number of generations", which corresponds to the number of iterations carried to tune the classifier, we set it to 15. An example of the best classifier for each feature type can be seen in Table ~\ref{featureTable}. Interestingly, decision tree-based algorithms and logistic regression outperformed others like SVMs. 

For our experiments, we extracted the 18 sound events from the two scenes using the annotations, and then we extracted different feature types from these isolated sounds. For each feature type experiment, the sound events' feature files were fed to Tpot in a randomly selected ratio of 75\% training and 25\% testing, each set with different files. We kept the same partitions across our experiments for consistency. The performance was measured in terms of accuracy and is displayed in Table ~\ref{featureTable}. 

\begin{table}[t]
\small
\centering
  \begin{tabular}{| l | c | c | }
    \hline
    \textbf{Feature Type} & \textbf{Accuracy}\% & \textbf{Classifier}  \\ \hline
    Home & 56.4 & Random Forest   \\\hline
    Home + G & 55.2 & Random Forest   \\\hline
    Home + G + P & 55.7 & Random Forest  \\\hline
    Residential & 53.3 & Gradient Boosting   \\\hline
    Residential + G & 57.8 & Decision Tree  \\\hline
    Residential + G + P & 56.7 & Random Forest  \\\hline
  \end{tabular}
\caption{Sound-event classification accuracy using the DCASE set up with four-folds partitions. The inclusion of the [G]eneric class improved performance for both scenes, whereas the inclusion of the [P]erturbed audio improved only the Home performance.}
\label{genericTable}
\vspace{-0.15in}
\end{table}
The features with the best performance were MFCCs with 67.7\% and thus we keep them for our DCASE evaluation set up. The other features have shown better results than MFCCs on audio classification, but it wasn't the case for this particular dataset. Results for Scatnet was 62.1\%, for GBF was 52.4\%, and for SGBF was 61.5\%. Moreover, the two normalized stacked features performed almost as good as MFCCs with 66.68\% for the stacked without PCA and 66.08\% for the stacked with PCA. In principle the stacked version contains more information about the acoustics and thus they were expected to perform better. Nevertheless, they didn't outperform MFCCs which is designed for speech and focus on lower frequencies rather than on a wider frequency range. We cannot draw a fundamental conclusion on the performance of these features for sound event classification since the amount of data and classes are determinant.

\begin{figure}[t]
   \centering
     \includegraphics[width=0.45\textwidth]{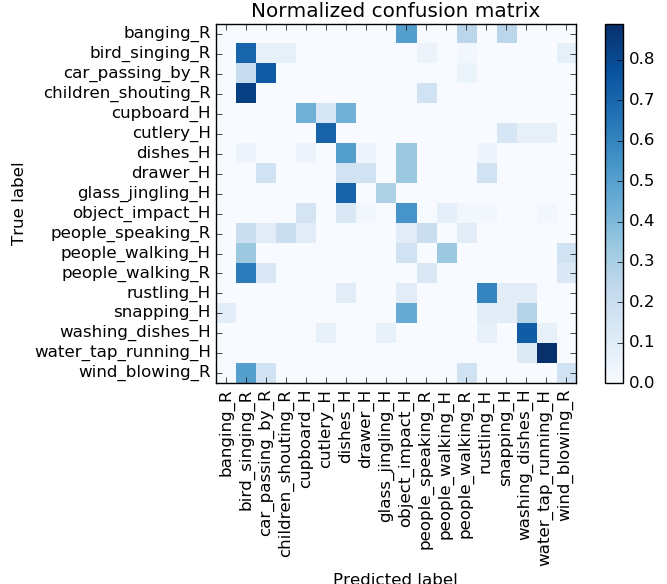}
     \caption{[H]ome and [R]esidential without the generic class. \emph{Object impact} and \emph{bird singing} capture most of ambiguities.}
     \label{fig:mfcc_features_18_normalized_confusion_matrix}
     \vspace{-0.25in}
\end{figure}
\vspace{-0.1in}

\subsection{Inclusion of Generic Sound Event Class}
\vspace{-0.1in}
\label{sec:generic}
In the annotated scenes, not every segment of audio corresponds to a labeled sound. Hence, it cannot be assured that any of our sound event classes have to be present on every test segment. To handle out-of-vocabulary segments, we proposed a generic sound event class. 

For the first experiment we wanted to analyze the impact of the generic class together with the 18 sounds in the multi-class classification set up described in Section~\ref{sec:features} using MFCCs. To create such class, we used the sound events annotations and trimmed out the audio between the labeled segments, which are unlabeled. Then, we randomly selected from both scenes, 60 audio files which is about the average number of sound event samples per class. In order to visualize the performance, we included the normalized confusion matrices (CMs) in Figures ~\ref{fig:mfcc_features_18_normalized_confusion_matrix} and ~\ref{fig:mfcc_features_18_plus_generic_normalized_matrix}. The accuracy performance without the generic class was 67.7\% and with the generic class was 60.94\%. The performance dropped with the inclusion of the new class, but we can also observe how although the generic class shared the background acoustics with the other sound classes, it didn't significantly ambiguate with them.

The second set of experiments used the DCASE setup of separate scenes and four folds, and utilized the sound events with and without the generic class, but this time the generic class will have files particular to the scene. The results can be seen in Table ~\ref{genericTable} showing benefit of including the generic class. Moreover, the CMs not included due to space limitations, had cleaner diagonals. The reasons for performance improvements on the DCASE set up are suggested by the utilization of less sound classes, which reduces class ambiguity. The utilization of the generic class built with same-scene files as opposed to a mix of both scenes. As well as the optimization per scene of the classifier using Tpot.
\vspace{-0.1in}
\subsection{Generation of Data Through Perturbation}
\vspace{-0.1in}
\label{sec:perturbation}
The scarcity of labeled data per event is a common issue as discussed in~\cite{kumar2016audio,anuragweakly}. Annotations are costly, sounds don't occur with the same frequency and in general it's hard to capture enough variations of the same sound to train robust models. To address this problem, multiple techniques have been explored in the literature such as perturbation of the audio signal as in~\cite{chen2015noise,kanda2013elastic}. The authors presented multiple types of perturbations resulting in improvements of speech separation. For Task 3 we performed time-based perturbation by speeding up and slowing down the sound event samples. We empirically analyzed multiple combinations of speed up-down values for different events. We concluded that speeding up more than 30\% the original signal resulted in unintelligible audio and speeding down the signal more than 100\% would be unlikely to occur. The range included 13 different speed values and the original version.
 
The set of experiments used the DCASE setup of separate scenes and four folds, and utilized the time-based perturbed audio. For training, we added to the original files the 13 versions of the perturbed audio files, whereas for testing, the set remained intact. The results can be seen in Table~\ref{genericTable}, where the performance for \textit{Home} improved, but not for \textit{Residential}. Thus, we decided to use perturbation for the DCASE evaluation.

\vspace{-0.1in}
\subsection{Sound Event Detection and Submission Systems}
\vspace{-0.1in}
\label{sec:sed}

For Task 3, we used the DCASE setup of separate scenes and four folds in a similar setup as the experiments from Table~\ref{genericTable}. For each scene, we extracted the sound events from the recordings using the annotations from the train set, followed by the extraction of MFCCs features. After, we trained the Tpot optimized multi-class classifier with the event samples. For testing, instead of using sound event files only, we segmented the scene recordings from the test set into one-second consecutive segments. This number was selected due to the metric schema of the DCASE evaluation, which considers one-second segments. After, we extracted audio features from the test segments and evaluate them with the classifier to obtain scores for each trained sound event class. The label corresponding to the highest score was chosen for the segments and then were written down into the DCASE format output file and fed to the official scoring scripts~\footnote{http://www.cs.tut.fi/sgn/arg/dcase2016/sound-event-detection-metrics} along with the ground truth to compute performance. 
\begin{figure}[t]
   \centering
     \includegraphics[width=0.45\textwidth]{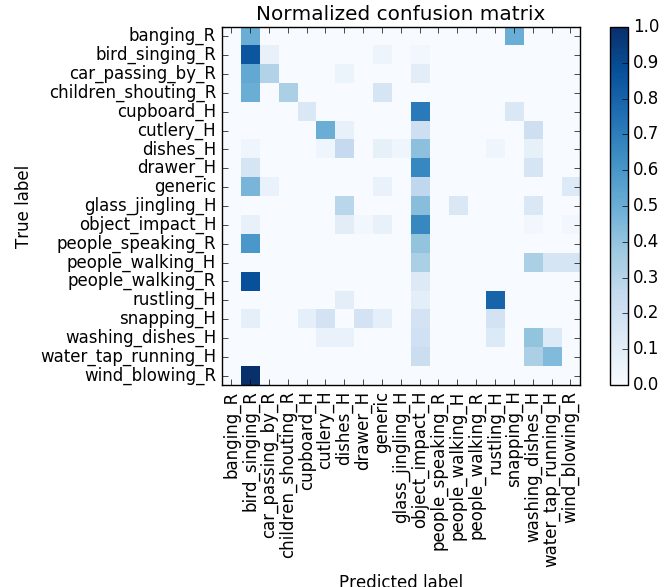}
     \caption{[H]ome and [R]esidential with the generic class. Although this class shared the background acoustics with the 18 sounds, it didn't cause major confusion.}
     \label{fig:mfcc_features_18_plus_generic_normalized_matrix}
     \vspace{-1em}
\end{figure}

We utilized the pipeline for three experiments, without generic class \& without perturbation, with generic class \& without perturbation and with generic class \& with perturbation. The results using the development-test set are shown in Table~\ref{SBER}. The inclusion of the generic class and the perturbation outperformed the baseline method by a significant margin for both \emph{Home} and \emph{Residential} scenes. Our submission consisted on the runs using G and G+P but using the evaluation set. The eval results were SBER of 0.9613 and Fscore of 33.6\% given by the G+P version.
\begin{table}[t]
\small
\centering
  \begin{tabular}{ | l | c | c |  }
    \hline
    \textbf{Acoustic Scene} & \textbf{SBER} & \textbf{F-score}\\ \hline 
    Home & 1.05 & 25.1 \\\hline
    Home + G & 0.91 & 23.7  \\\hline
    Home + G + P & 0.9 & 24.7   \\\hline
    Residential & 0.64 & 54.6 \\\hline
    Residential + G & 0.72 & 45.9   \\\hline
    Residential + G + P & 0.63 & 52.2   \\\hline
  \end{tabular}
\caption{Our Segment-based Error Rate, using [G]eneric and [P]erturbation, outperformed the baseline.}
\label{SBER}
\vspace{-1em}
\end{table}
\vspace{-0.15in}

\section{Conclusion}
\vspace{-0.1in}
\label{sec:conc}
In this paper we showed different approaches for both acoustic scene classification (Task 1) and sound event detection (Task 3) of the 2016 DCASE challenge. On both tasks we were able to obtain significant improvement over the baseline method. For Task 1 we observed that the $\vec{\beta}$ features performed much better than $\vec{\alpha}$ features. Although, linear and RBF kernels with $\vec{\beta}$ features can outperform the baseline by considerable margin on its own, we make note of the fact that a multiple classifier system can give further improvements. For Task 3, we tested different features and classifiers and significantly improved the baseline. Moreover, we explored a way of handling out-of-vocabulary sound segments with the generic class and the inclusion of perturbed audio to add robustness.
  
\pagebreak
\ninept
\bibliographystyle{IEEEtran}
\bibliography{refs}

\end{sloppy}
\end{document}